\documentclass[aps,pra,reprint,superscriptaddress,showpacs,showkeys]{revtex4-2}

\usepackage[multidot]{grffile}
\usepackage{amsmath,amssymb}  
\usepackage{color}
\usepackage{graphicx}

\usepackage{titlesec}
\usepackage{etoolbox}

\makeatletter
\patchcmd{\ttlh@hang}{\parindent\z@}{\parindent\z@\leavevmode}{}{}
\patchcmd{\ttlh@hang}{\noindent}{}{}{}
\makeatother
\titleformat*{\section}{\normalsize\bfseries\filcenter}
\titleformat*{\subsection}{\normalsize\bfseries\filcenter}
\titleformat*{\subsubsection}{\normalsize\itshape\filcenter}

\draft % marks overfull lines with a black rule on the right

\begin{document}

%\preprint{}

\title{Phase-incoherent photonic molecules in V-shaped mode-locked
	vertical-external-cavity surface-emitting semiconductor lasers} %Title of paper


\author{Jan Hausen}
\email[]{hausen@campus.tu-berlin.de}
%\thanks{}

\affiliation{ Institute of Theoretical Physics, Technische Universit{\"a}t Berlin, Hardenbergstraße 36, 10623 Berlin, Germany}

\author{Stefan Meinecke}
\email[]{meinecke@tu-berlin.de}
%\thanks{}
\affiliation{ Institute of Theoretical Physics, Technische Universit{\"a}t Berlin, Hardenbergstraße 36, 10623 Berlin, Germany}

\author{Julien Javaloyes}
%\thanks{}
\affiliation{Departament de F\'{\i}sica, Universitat de les Illes Balears \&
Institute of Applied Computing and Community Code (IAC-3), Cra.\,\,de
Valldemossa, km 7.5, E-07122 Palma de Mallorca, Spain}
\author{Svetlana V. Gurevich}
\affiliation{Departament de F\'{\i}sica, Universitat de les Illes Balears \&
Institute of Applied Computing and Community Code (IAC-3), Cra.\,\,de
Valldemossa, km 7.5, E-07122 Palma de Mallorca, Spain}
\affiliation{Institute for Theoretical Physics, University of M{\"u}nster, Wilhelm-Klemm-Str. 9,
48149 M{\"u}nster, Germany}
%\thanks{}


\author{Kathy L{\"u}dge}
\email[]{kathy.luedge@tu-berlin.de}
%\thanks{}
\affiliation{ Institute of Theoretical Physics, Technische Universit{\"a}t Berlin, Hardenbergstraße 36, 10623 Berlin, Germany}

\date{\today}

\begin{abstract}
Passively mode-locked vertical external-cavity surface-emitting semiconductor lasers (VECSELs) composed of a gain chip and a semiconductor saturable absorber have been drawing much attention due to their excellent performance figures. In this work we investigate  how localized structures and incoherent, non-locally bound, pulse molecules emerge in a long cavity VECSELs using a V-shaped cavity geometry.  We show that these states are bistable with the laser off state and that they are individually addressable.
Using a model based upon delay differential equations, we demonstrate that pulse clusters result from the cavity geometry and from the non-local coupling with the gain medium; in the long cavity regime this leads to locally independent, yet globally bound phase, incoherent photonic molecules. Using a multiple time-scale analysis, we derive an amplitude equation for the field that allows us to predict analytically the distance between the elements of a cluster.
\end{abstract}

\maketitle %\maketitle must follow title, authors, abstract and \pacs

\section{INTRODUCTION}
\label{I}

Vertical external-cavity surface-emitting semiconductor lasers (VECSELs)
composed of a gain chip and of a semiconductor saturable absorber mirror (SESAM) are an important class of passively mode-locked (PML) lasers
\cite{haus00rev}. These devices are able to generate transform limited
pulses in the 100~fs range with peak powers of 500~W \cite{WAL16}
at GHz repetition rates. In the limit of cavity round-trips much longer than 
the gain recovery time, typically a nanosecond in semiconductor materials, 
the mode-locked pulses found
in VECSELs may coexist with the off solution. In these conditions,
they can be interpreted as \emph{temporal localized structures} (LSs)
\cite{MAR14c}. This regime can be leveraged to generate PML states 
composed of individually addressable pulses at arbitrary low repetition rates,
using, e.g., current or optical modulation~\cite{CAM16,JAN15,MAR15b,KUR19}.

Localized structures are ubiquitous in dissipative systems and they have been widely observed in nature \cite{WU84,MOS87,FAU90,PIC91,NIE92,VAN94,LEE94,UMB96,AST01},
as well as in a variety of driven photonic non-linear systems such as ring resonators \cite{JAN15,GUS15}, micro-cavities \cite{YI18} and external cavities fed by a vertical cavity surface emitting lasers \cite{BTB-NAT-02,TAN08b}. Due to the extension of the original concept
of conservative solitons, LSs are also sometimes referred to as dissipative solitons \cite{AKH05,GRE12b}. 
In photonic systems LSs occur either transversely to the propagation direction \cite{LUG03,ACK09,MAR15d} or in the longitudinal (temporal) direction \cite{BAR17a,HER14b,MAR14c,MAR15b}. Localization in both 
directions would potentially lead to light bullets \cite{BRA04,PIM13,JAV16,GUR17}. 

This paper focuses on the dynamics of temporal LSs found in VECSELs arranged in 
a V-shaped cavity configuration~\cite{WAL16,WAL18,HAU19,GRO20}. At variance with previous
works that employed linear cavities, the gain chip in the case of the V-shaped  configuration is positioned between the absorber and the output mirror. 
In addition to the LSs that also exist in linear cavities, the V-shaped
geometry induces phase-incoherent photonic molecules, i.e. pulses which are globally 
bound but locally independent. Their temporal separation can be controlled 
tuning the difference of optical path between the two arms of the ``V'' as
pulses pass through the gain section twice per round-trip inducing the 
necessary non-local influence. In the long cavity regime, these can 
be referred to as weakly bound ``catenane molecules'' like those found 
recently in \cite{MAR15d,JAV17,GUI18}. Similar molecular soliton structures have also been examined using Kerr cavities \cite{WEN20}, mode-locked fibre lasers \cite{ORT10,ZAV09,WAN19e} or optical fibres pumped by solid state lasers \cite{STR05b, HAU08, KRU17b,HER17a,KUR19}.

Temporal localization in photonic systems can be taken advantage of to realize a variety of applications \cite{GRE12b}. In particular, LSs can be used as elementary bits of information while the cavity acts as an all-optical buffer \cite{GEN08,LEO10,MAR14c,GAR15,PAN16e}. With regard to optical communications, soliton molecules have been suggested to extend the binary alphabet to enhance optical data transfer by transferring several bits simultaneously to circumvent the Shannon limit \cite{ROH12a,YUS14,MIT16}.
Individual addressing opens further interesting possibilities for the optical generation of arbitrary trains of spikes, which has potential applications in different domains.
Due to the observed incoherence between pulses inside a single molecule they could be applied in future spectroscopy applications demanding dense frequency combs but a variable distribution of the comb power \cite{SPH-NAP-12,CHE10d,COI14,TLI17}. Here, since the various pulses do not generate a coherent beating, the
comb would not be spectrally modulated. The intensity of the teeth in the comb would scale with the number of pulses, while avoiding the resolution decreases incured by having N pulses in the cavity usually leading to a N-fold increase of the comb spacing. This might also find applications in pump-probe sensing of material properties \cite{PED08}. Furthermore, the possibility to create specific, stable pulse patterns from a single device can also be of use for material processing \cite{KER16}. \\ 
As mode-locked VECSEL optical systems have a wide array of applications further understanding their dynamics as a function of the cavity parameters to improve the performance or prevent certain dynamical regimes is essentially a topic relevant to applied physics.\\
As such, the possibility to generate controllable cluster of LSs using non-local effects has both fundamental and practical interests.

\section{EMERGENCE OF LSs}
\label{II}
\begin{figure}[t]
	\includegraphics{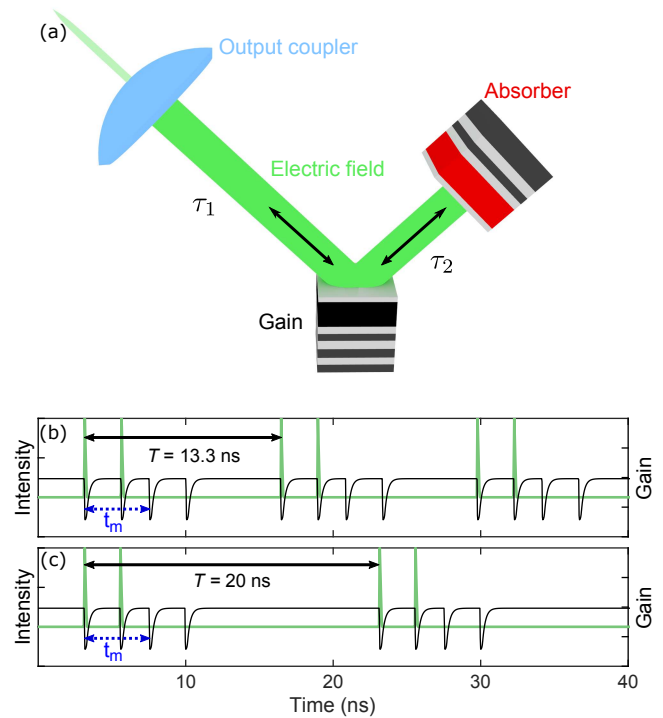}
	\caption{Setup of a passively mode-locked VECSEL with V-shaped cavity geometry. The main constituents are a semiconductor saturable absorber mirror, an out-coupling facet with high reflectivity ($\kappa \approx 0.99$) and a semiconductor gain chip which can be optically or electrically pumped. The time of flight in the two cavity arms are $\tau_1$ and $\tau_2$, respectively. (b)-(c) Photonic molecules in the long cavity limit for fixed $\tau_2 = 2.2$\,ns but at varied cavity length $T$, with $t_m$ referring to the size of the molecule.} 
	\label{Fig1}
\end{figure}

The setup of V-shaped passively mode-locked VECSEL is sketched in Fig.~\ref{Fig1}. There, the gain chip is placed in a central position while the semiconductor saturable absorber and the output coupler are located on each side of the cavity. A single pulse depletes the gain twice during one round-trip as it propagates forward and then backward through the cavity (black arrows in Fig.~\ref{Fig1}(a)). The cavity configuration can be described by the length of the two cavity arms $\tau_1=L_1/c$ and $\tau_2 = L_2/c$. 

\begin{figure}[t]
	\includegraphics{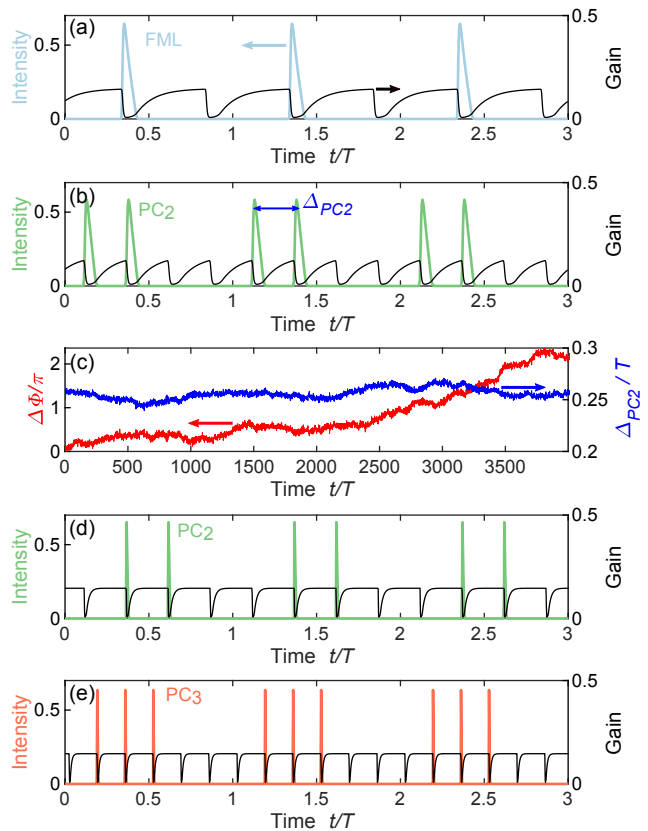}
	\caption{Localized structures (coloured pulses) and the corresponding evolution of the gain G (black lines) in the intermediate and long symmetric cavity regime found by direct numerical integration of Eqs.~\eqref{Eq:E}-\eqref{Eq:R}. 
(a) Single pulse regime, a secondary ghost gain depletion corresponding to the second gain pass is clearly visible. (b) Pulse cluster with two elements (PC$_2$). (c) Phase difference between the two elements of the PC$_2$ bound state $\Delta \phi$ (in red). It is freely drifting while the relative distance ($\Delta_{PC_2}$ in blue) between the two bits remains constant, up to the influence of noise. In all cases presented in (a,b,c) $T =1.875$\,ns. (d) Evolution of the PC$_2$ regime in the long cavity regime $T =12.5$\,ns and (e) PC$_3$ dynamics with $T =25$\,ns.
Other parameters are $\gamma_g = 5$\,ns$^{-1}$, $\gamma_q = 180$\,ns$^{-1}$, $Q_0 = 0.177$, $\gamma = 240 $\,ns$^{-1}$, $G_0/G_{th} = 0.8$, $\kappa = 0.99$, $s = 20$ and $\alpha_{g,q} =0$. For (c) $\alpha_g = 1.5$, $\alpha_q = 0.5$ and the noise is added as $\sqrt{\frac{\eta}{dt}}D$ in each integration step for Eq.~\eqref{Eq:E}, with Gaussian white noise term $D$ and the noise strength $\eta = 25$ (normalized to $T$).} 
	\label{Fig2}
\end{figure}

As in the experimental realization in~\cite{WAL16,WAL18}, we model our system with the delay differential equations (DDEs) derived in Appendix A of~\cite{HAU19} and based upon~\cite{VLA09}. The theoretical model yields the dynamics of the electric field $E$ and the integrated carrier densities of the saturable absorber $Q$ and of the gain $G$ as follows:  
\begin{align}
\label{Eq:E} \dot{E} =& -\gamma E+\gamma E(t-T)R(t-T),\\
\label{Eq:G} \dot{G} =& \gamma_g ( G_0-{G})-(e^{G}-1)\times\nonumber\\
&\{\mid {E}\mid^2+\mid E(t-2\tau_2)\mid^2e^{G(t-2\tau_2)-2Q(t-2\tau_2)}\},\\
\label{Eq:Q} \dot{Q} =& \gamma_q( 
Q_0-Q)+s(e^{-2Q}-1)e^{G}\mid E\mid^2,\\
\label{Eq:R}
R(t) =& \sqrt{\kappa}e^{\frac{1-i\alpha_g}{2}[G(t+2\tau_2)+G(t)]-(1-i\alpha_q)Q(t)},
\end{align}
where $\gamma$ describes the resonance width of the gain chip, $\gamma_{g,q}^{-1}$ are the carrier life times in the gain and the absorber section, $G_0$ is the unsaturated gain driven by the pump power, $Q_0$ is the unsaturated absorption in the absorber chip, $\kappa$ models the non-resonant losses per round-trip, the factor $r_s$ is proportional to the ratio of the differential gain coefficients and confinement factors in the two active sections and $\alpha_{g,q}$ are the line-width enhancement factors in the gain and absorber section, respectively.
The algebraic equation (\ref{Eq:R}) describes the total amplification and losses of the electric field during one round-trip in the laser cavity. We note that the cold cavity round-trip time is $T=2(\tau_1+\tau_2)$, see Fig.~\ref{Fig1}, such that $\tau_1 =\tau_2 = 0.25T$ describes the symmetric V-shaped cavity. We note that we have applied a coordinate transformation of the time $t$, to reduce the number of delayed terms (see detailed explanation in Appendix \ref{AppTrafo}).
\ \\
We use a set of standard parameters depicted in the caption of Fig.~\ref{Fig2} to analyze the system in the long cavity regime below the lasing threshold, i.e., we set $G_0$ such that $G_0<G_{th}$, where $G_{th}=Q_0-\dfrac{1}{2}\log{\kappa}$ is the continuous wave lasing threshold. This means that single pulses can be individually excited while being multi stable with the homogeneous "off" solution \cite{BAR17a,HER14b,MAR14c,MAR15b}. At this point we emphasize that the general appearance of a localized structure is not affected by the amplitude phase coupling and therefore we set $\alpha_{g,q} = 0$ in the first steps of our study \cite{MAR14c}.\\
In general, pulses interact with each other via the overlap of their tails and, at long range, the slowest variable to relax is the gain material one, i.e. $G(t)$ is relaxing on the time-scale of $\gamma_g^{-1}$. Repulsive interactions between pulses stem from the gain exponential tails \cite{CAM16,KUT98} thereby favoring an equidistant pulse spacing, which leads to harmonic mode-locking (HML). The relaxation towards equidistant pulses in the HML solutions can be arrested by using a sufficiently long cavity. More precisely, the transitory time can be made exponentially large. Therefore, we set $\gamma_g T \gg 1$. Due to the double gain pass of a single pulse during each round-trip in the V-shaped cavity (Fig.~\ref{Fig1}(a)), additional pulse depletions appear. In the long cavity limit a second pulse can be trapped in in-between the two gain depletions of the preceding pulse as indicated by the blue dashed arrows in Fig.~\ref{Fig1}(b-c). In this molecule structure (photonic molecule), the second pulse is locally independent as the gain completely relaxes between the passes of the two pulses, but globally bound by the second gain depletion of the preceding pulse. In this situation the size of the molecule $t_m$ is only dependent on the smaller delay $\min(\tau_2,\tau_1)$ and not on the size of the complete resonator $T$ as indicated in Fig.~\ref{Fig1}(b-c); it is because $\min(\tau_2,\tau_1)$ governs the timing of the second gain depletion.
\ \\
Similarly to the LSs investigated in face-to-face coupled cavities \cite{MAR14c,MAR15b,CAM16}, we find that LSs can be triggered in sub-threshold V-shaped cavities, and that the latter are multi-stable to the "off" solution, see Fig.~\ref{Fig2}(a) and the appendix \ref{AppA}. However, a unique characteristic of the V-shaped cavity, i.e. a second gain pass per round-trip, leads to a secondary gain depletion as observed in Fig.~\ref{Fig2}(a) that seems to be triggered by a "ghost" pulse (see Appendix \ref{AppTrafo}). The gain temporal profile in Fig.~\ref{Fig2}(a) highlights how a single pulse can induce (apparent) non-local effects in the active material as pointed out in \cite{WAL18,HAU19}.
In addition, the non-local gain dynamics allows the creation of pulse cluster of $n$ pulses (denoted PC$_n$). An example of such a regime is depicted in Fig.~\ref{Fig2}(b) in the intermediate cavity regime ($T = 1.875$\,ns); as each pulse creates two depletions of the gain (shown in black), the latter is not able to fully recover to the equilibrium; therefore the pulses within a cluster are strongly bound in this case. This can be asserted by representing the behavior of their relative distance $\Delta_{PC_2}$, see blue line in Fig.~\ref{Fig2}(c); while it fluctuates around an equilibrium value of $\Delta_{PC_2} \simeq 0.25T$, it does not converge to that of an harmonic mode-locking solution $\Delta_{HML_n} \simeq T/n$ \cite{KUT98}. However, the relative phase between the elements of a cluster is free to drift, see red line in Fig.~\ref{Fig2}(c). Averaging over different noise realizations, we find that the ensemble variance of the phase grows linearly with time, revealing a diffusive behavior. These results convincingly prove that the LSs forming clusters are not phase coherent one with each other and therefore can be identified as phase incoherent photonic moleculess.

The PCs solutions persist in a long cavity regime as shown in Fig.~\ref{Fig2}(d,e). Here the gain relaxes to its equilibrium in-between pulses. This leads to a scenario in which the second pulse is locally independent from the first one due to the flat gain landscape between the first pulse and its ghost image (second gain pass) in the middle of the round-trip. Yet, the second pulse cannot approach the ghost replica as it would encounter a depleted gain, which generates repulsive forces. As such, the second pulse is globally bound due to the non-local dynamics of the gain; such a mechanism is similar to that of the nested molecules found in~\cite{MAR15d,JAV17,WAN19e}. In this case the size of the molecule is only dependent on the distance between the two gain passes of one pulse.
At even higher round-trip times ($T = 25$\,ns), similar higher order locally independent yet globally bound PCs can be observed; the PC$_3$ solution is shown in Fig.\ref{Fig2}(e). Here also, although the pulses are bound within a cluster, they are mutually incoherent. Notice that multi-peak bound structures also exist in ring mode-locked cavities, but only as unstable solutions and with varying pulse amplitudes among the structures \cite{SCH18e} or induced by a periodic excitation scheme\cite{MAR15b}. 

\begin{figure}[b]
	\includegraphics{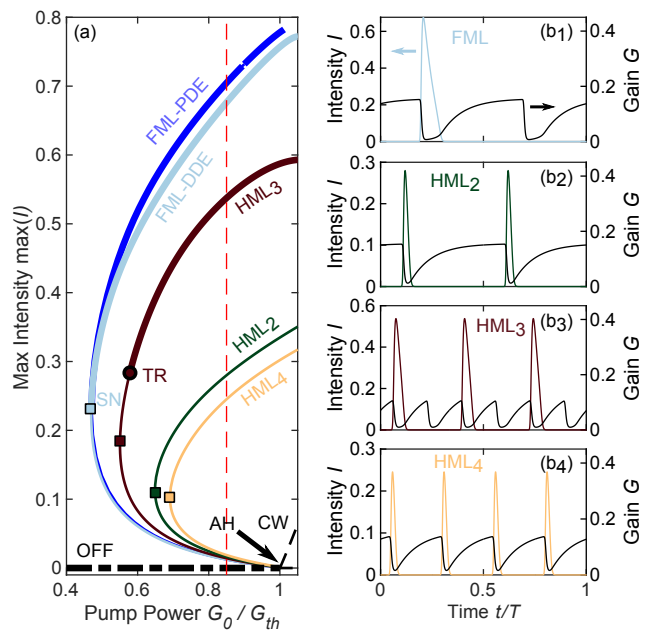}
	\caption{(a) Evolution of the maximum pulse intensity with respect to the normalized pump power $G_0/G_{th}$ of the fundamental (FML) and harmonic (HML$_n$) mode-locking solutions, where subscript $n$ denotes the number of equally spaced pulses. Thick and thin lines indicate stable and unstable dynamics, respectively. The solutions are born in subsequent Andronov Hopf bifurcations (AH) along the steady state CW branch (intensity stretched by a factor of 40). The FML branch is found utilizing the path continuation of the DDE model~\eqref{Eq:E}-\eqref{Eq:R} (light blue) and the Haus master equation~\eqref{EHaus} (dark blue). The branches all fold back in saddle-node bifurcations (squares) and become multi stable to the off solution depending on the cavity geometry. The HML$_3$ solution stabilizes in a torus bifurcation TR (circle). (b$_{1-4}$) Gain and field intensity profiles at $G_0/G_{th} = 0.85$ (marked by dashed line in (a)) and symmetric cavity $\tau_1 = \tau_2 = 0.25T$. Other parameters as in Fig.\ref{Fig2}(a).} 
	\label{Fig3}
\end{figure}

Utilizing the path continuation software DDE-Biftool~\cite{DDEBT}, we can gain a deeper insight into the behavior and dynamics of the harmonic mode-locking solutions and pulse clusters in/near the localized regime. The maximum intensity for the different branches of harmonic solutions as a function of the pump power $G_0$ are represented Fig.~\ref{Fig3}(a), while the corresponding temporal outputs are depicted in Fig.~\ref{Fig3}(b$_{1-4}$) in matching color. Similar to the results obtained for face-to-face coupled cavities \cite{MAR14c,MAR15b,SCH18e}, equally spaced temporal LSs are generated via Andronov-Hopf (AH) bifurcations along the continuous wave (CW) branch continued in pump power $G_0$. 
The fundamental (FML) and harmonic mode-locking solution branches HML$_n$, distinguished by the number $n$ of pulses in the cavity, emerge subcritically; it means that they appear as unstable solution branches, see thin lines in Fig.~\ref{Fig3}(a). The branches then fold at a saddle node SN bifurcation point (labeled with squares in Fig.~\ref{Fig3}(a)). Further, they may give rise to a stable solution, see thick lines in Fig.~\ref{Fig3}(a). It is by this mechanism that the various regime exist below the CW threshold, i.e. $G_0 < G_{th}$. All these solutions branches are therefore multi-stable with the off-solution, which is an essential criterion for the pulses to become localized \cite{GRE12b,MAR14c}. 

\ \\
The FML solution stabilizes at the SN point. However, the HML solutions become stable in torus bifurcations TR (labeled with circles) occurring at pump currents slightly above the respective SN points (see Fig.~\ref{Fig3}(a)). This stabilization mechanism is different from the case of a face-to-face coupled cavity, where all solutions would stabilize at their respective SN points~\cite{MAR14c,SCH18e}. 
In the symmetric cavity configuration chosen in Fig.\ref{Fig3}, the regular temporal separation between pulses in the harmonic solutions HML$_2$ and HML$_4$ make it so that one can find only as many gain depletions as there are pulses, and not twice as much as for the HML$_3$ branch. In the former case each gain depletion results from the combined effect of the influence of a pulse and the ghost of another. In other words the emission of the second pulse coincides with the backpropagation of the first pulse through the gain. This is also the reason why the HML$_2$ and HML$_4$ solutions do not re-stabilize in Fig.~\ref{Fig3}(a). However, re-stabilization can be obtained for asymmetrical cavity configurations, which will be discussed further.
\begin{figure}
	\includegraphics{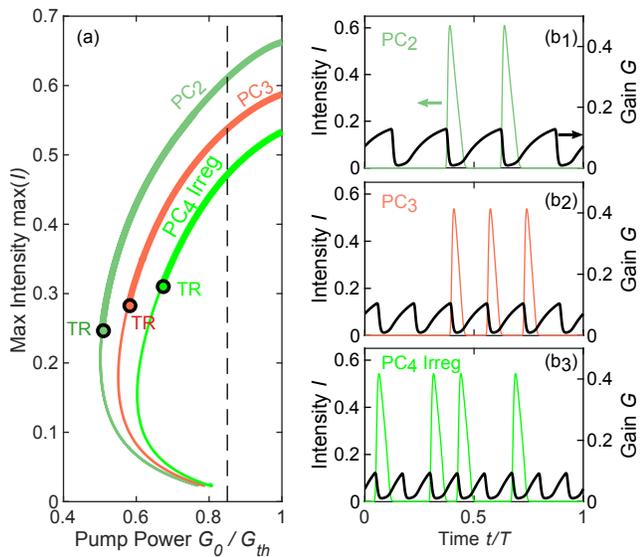}
	\caption{(a) Maximum pulse intensity of bound PC states with different numbers of pulses with respect to the normalized pump power $G_0/G_{th}$. Thick and thin lines indicate stable and unstable solutions, respectively. The branches of the PC solutions with two and three pulses (PC$_2$ and PC$_3$) as well as the irregular four pulses solution (PC$_4$ Irreg) all become stable below the threshold $G_{th}$. The stabilizing torus bifurcations (TR) are marked by black circles. The corresponding electric field intensity and gain profiles in the stable regions at $G_0 = 0.85 G_{th}$ (dashed line) are displayed in (b$_{1-3}$). The PC solutions are shown for a symmetric cavity $\tau_2 = 0.25T$, all other parameters are as given in Fig.\ref{Fig2}(a).} 
	\label{Fig4}
\end{figure}

We now turn our attention to the pulse clusters observed in Fig.~\ref{Fig2}. The bifurcation diagram of the PC$_2$ and PC$_3$ solutions is shown in Fig.~\ref{Fig4} for an intermediate cavity length ($\gamma_g T = 9.4$). We notice that although the clusters are composed of non-equidistant pulses, see Fig.~\ref{Fig4}(b$_{1-2}$), their separation is such that, considering the additional ghost pulses due to non-local effects, they induce equi-spaced carrier gain depletions. One understands that this leads to the larger gain extraction and to the more energetically favorable situation. In addition, it is also possible to find "irregular" pulse cluster solutions as indicated by the PC$_4$ Irreg. solution in Fig.~\ref{Fig4}b$_{3}$. In order to understand these "irregular" regimes, one simply has  to make a permutation between a real pulse and its ghost, thereby generating the same temporal profile for the gain dynamics. An example is shown in black in Fig.~\ref{Fig4}b$_{3}$. A similar irregular regime can be found for the PC$_3$ cluster.
The PC states are multi stable below the threshold in the same regions as the harmonic mode-locking solutions as discussed before in Fig.\ref{Fig3}(a). Equally, they stabilize in torus bifurcations (circles labeled with TR in Fig.\ref{Fig4}(a)) slightly above the SN point, i.e. the lower existence boundary. Following the PC$_2$ solution down to small pulse intensities, we find that it is born in a period doubling bifurcation (PD) of the HML$_2$ solution. Along the solution branch the two pulses of the PC$_2$ adjust their relative temporal distance from $\Delta_{PC_2} = 0.5T$ at the PD point (a zoom of this region can be found in  see Fig.\ref{Fig9} in Appendix \ref{APPPC2}) to $\Delta_{PC_2} = 0.25T$ (PC$_2$) in the stable regime. It has to be noted that the generation mechanism of the PC solutions is different from that observed in the non-localized mode-locking regime. In the latter, additional pulses continuously emerge increasing the bias along the FML solution branch \cite{HAU19}. \\ 
The position of a solitary pulse in a PC is always limited by the two surrounding gain depletions, see e.g., Fig.~\ref{Fig2}(d$_2$). We can assess the nature of this bond in dependence of the cavity round-trip time by performing a Floquet analysis~\cite{KLA08} of the periodic PC$_n$ solutions as described in~\cite{MAR15d}. 
In particular, for intermediate round-trip times where the gain does not relax to the equilibrium value only one neutral mode exists for the pulse cluster in the cavity (see Fig.~\ref{Fig11} in Appendix \ref{APPFloquet}), i.e. one Floquet multiplier is located at $\mu = 1$. This multiplier corresponds to the neutral mode of translation of the entire waveform. It is present in any dynamical system without explicit time dependence. As a counter-example, this multiplier would disappear in the analysis of an actively mode-locked laser because the pulse timing is locked to that of the externally imposed modulation. In our case, as the round-trip time is increased, $n-1$ additional Floquet multipliers approach $\mu = 1$, where $n$ is the number of pulses in the PC solution. As the gain fully relaxes in-between pulses, $n$ Floquet multipliers at $\mu=1$ can be found as shown for the PC$_2$ solution in Fig.~\ref{FigFloquet} in Appendix \ref{APPFloquet}. Hence, there is one neutral mode of translation for each pulse in one cluster which demonstrates that they are locally independent. Notwithstanding, they remain globally bound since for each pulse, the previous and next gain depletions induce repulsive forces. These incoherent molecules are similar to the nested molecules found in \cite{MAR15d,JAV17}. As the Floquet exponents give a measure of the residual interaction of the pulses, one can note that the transition between locally independent molecules and bound states is continuous depending on the round-trip time (see exponential behavior in Fig.~\ref{FigFloquet}(f) in Appendix \ref{APPFloquet}). 

\begin{figure}
	\includegraphics{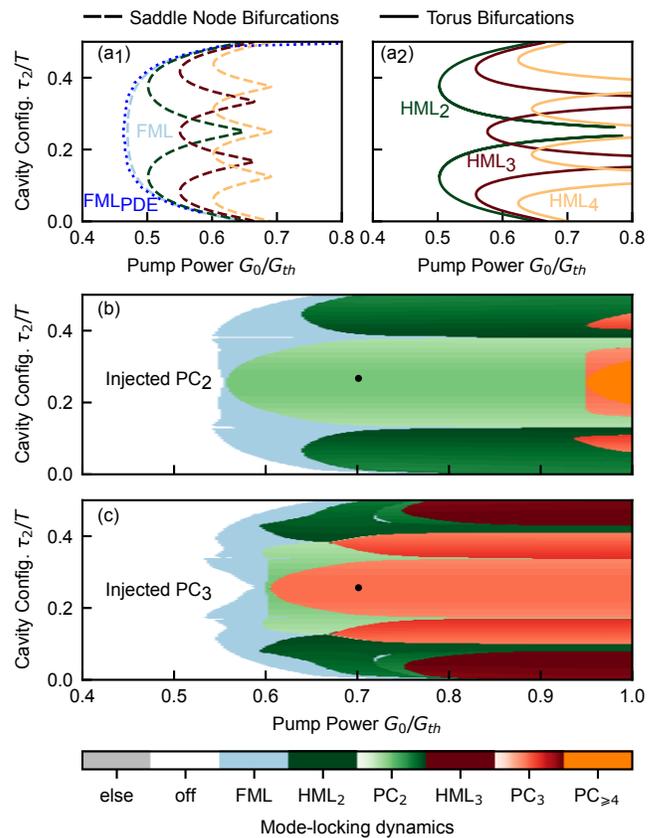}
	\caption{(a$_1$) Saddle-Node (folding) and (a$_2$) torus (stabilizing) bifurcations in the $(G_0,\tau_1)$ plane; the colors indicate fundamental mode-locking (FML), harmonic mode-locking (HML) and pulse cluster (PC) dynamics. Subscripts label the number of pulses in the cavity, $\alpha_{g,q} = 0$. For comparison the numerically found saddle-node line of the FML solution resulting from the PDE model is plotted in dark blue in (a$_1$). (b)-(c) Stable localized pulsations, evolving after injecting different PC solutions into the empty cavity as initial condition (stable at the black dots). The injected solutions were found to be stable after $10^4$ round-trips at $G_0 = 0.7 G_{th}$. The injected solutions are: (b) pulse clusters with two pulses (PC$_2$) and (c) three pulses (PC$_3$). The color coding is such that pulse distance $\Delta_{PC_n}$ increases for lighter shading. $(\alpha_g,\alpha_q) = (1.5,0.5)$, all other parameters as given in Fig.~\ref{Fig2}(a). } 
	\label{Fig5}
\end{figure}

In order to investigate the effect of the cavity geometry on the existence of different localized multi-pulse states, we explored how the stabilizing torus and saddle-node SN bifurcations of the HML and FML solutions in Fig.~\ref{Fig3}(a)) change as a function of the pump power and of the cavity asymmetry. 
By introducing the condition $2\tau_1 + 2\tau_2 = T$ we keep the cavity round-trip time $T$ constant, while changing the cavity geometry. The resulting SN and TR lines in the ($G_0/G_{th}$,\,$\tau_2/T$)-plane are shown in Fig.~\ref{Fig5}a$_{1-2}$, respectively. 
Both bifurcations clearly show a resonance behavior with respect to the position of the gain chip $\tau_2$. For certain cavity configurations no stabilizing torus bifurcation can be found and the curves in Fig.~\ref{Fig5}(a$_2$) are discontinuous; at these parameter values, two pulses collide in the gain chip as the distance between the pulses equals the propagation time between the two gain passes (e.g. $2\tau_2 = 0.5T$ for HML$_2$ or $2\tau_2 = 0.33T$ for HML$_3$). Hence, the number of resonances equals the number of pulses in the corresponding HML solution. This behavior can also be noted in Fig.\ref{Fig3}(a) as there is no stable region for the even HML$_2$ and HML$_4$ states at $\tau_2=0.25T$. \\
Oppositely, resonances at which the stabilizing torus and lower saddle node reach the lowest pump power can be found in Fig.~\ref{Fig5}(a$_{1-2}$). 
They correspond to the cavity configurations at which the maximum gain saturation can be achieved due to the pulses-passes happening at the largest temporal interval which is why the largest temporal distance is favorable. Maximizing gain extraction further explains why, in a symmetric cavity, the PC$_2$ solution is stable while the HML$_2$ is not, cf. Fig.~\ref{Fig3}(b$_2$) and Fig.~\ref{Fig4}(b$_1$); the former induces four gain depletion while the latter creates only two. \\
In order to obtain an overview of the stable regions of the PC states, we injected a PC solution found at a symmetric cavity configuration with $G_0 = 0.7 G_{th}$, see the black dots in Fig.\ref{Fig5}(b,c), and record to which solution the system converges after an integration time of $10^4$ round-trips. This analysis was performed in the ($G_0/G_{th},\,\tau_2/T$)-plane. We set $(\alpha_g,\alpha_q) = (1.5,0.5)$ to better match previous works \cite{CAM16,SCH18e,CAM18}. Our results are shown in Fig.~\ref{Fig5}(b,c). The color-code distinguishes the localized solution to which the system relaxes after a transitory regime. \\
To ensure optimal gain depletions, the pulse distances in the PC solutions are additional degrees of freedom by which these states can adapt to changes in the length of the cavity arms. The variation of the pulse distances in the stabilized PC solution is encoded in the shading of the colors indicating PC$_2$ and PC$_3$ solutions in Fig.~\ref{Fig5}(b,c), with ascending pulse distance from bright to dark colors. If the cavity is shifted slightly away from a symmetric positioning, say $\tau_2 = 0.25T$ in Fig.~\ref{Fig5}(b,c) for the PC$_2$ solution, the distance decrease to adjust for an optimal gain depletion. In case the gain chip is placed at one of the edges of the cavity, top and bottom in Fig.~\ref{Fig5}(b,c), a solution with (almost) equally spaced pulses stabilizes. Hence, incoherent photonic molecules with specific pulse distances can be designed by adequately tuning the cavity geometry. 

\section{NON-LOCAL HAUS MASTER EQUATION}
\label{III}
In order to understand the relationship between pulse distances and cavity configuration, we derive a Haus master equation model for the investigated V-shaped laser system~\eqref{Eq:E}-\eqref{Eq:R}. Using a decomposition of the dynamics in slow and fast stage, one can obtain an analytical expression for the pulse distance and the pulse power as a function of $\tau_1$,$\tau_2$.  

We start the derivation of a Haus master equation model \cite{HAU00} from the DDEs~\eqref{Eq:E}-\eqref{Eq:R} describing the V-shaped cavity system. The resulting partial differential equation (PDE) system could potentially also be used to study dispersive effects and spatio-temporal instabilities \cite{VLA14,JAV16,GUR17}. Although it is possible to investigate dispersion effects using a DDE approach \cite{PIM17,SCH19b}, it is a more demanding and less intuitive approach. The details of derivation of the master equation model is given in Appendix~\ref{APPHaus}; it is based on the multi-time scale analysis presented in \cite{KOL06}, but can in principle also be done utilizing the functional mapping approach outlined in \cite{SCH18f,SCH20d}. For the multi-time scale analysis, we introduce the slow time-scale $\theta$ corresponding to the electric field evolution from round-trip to round-trip and the fast time-scale $\sigma$ describing the pulse shape within one round-trip \cite{CAM16,VLA05}. Furthermore, we assume the limit of small gain $G$, which is fulfilled in the sub-threshold regime as $G_{th} = 0.18$ for the chosen parameters and also the absorption is small as $Q_0 = 0.18$. Hence, we are investigating the system in the so-called uniform field limit, which refers to the gain, absorption, losses ($1-\kappa=0.01$), and spectral filtering being small. The full partial differential equation (PDE) system then reads:
\begin{align}
\label{EHaus}
\partial_{\theta} E &=\frac{1}{2 \gamma^2} \partial^2_{\sigma} E +\biggl[\frac{1}{2}(1-i\alpha_g) \left(G(\sigma-\frac{T-2\tau_2}{T})+G\right)- \nonumber \\ & 
(1-i\alpha_q)Q+\frac{1}{2}\log(\kappa)\biggr]E,\\
\partial_{\sigma} G &= \gamma_g G_0-\gamma_g G-G\left( |{E}|^2+|E(\sigma-\frac{2\tau_2}{T})|^2\right),\\ 
\label{QHaus}
\partial_{\sigma} Q &=  \gamma_q Q_0-\gamma_qQ-2s Q|E|^2,
\end{align}
where all parameters have the same meaning as in the DDE system~\eqref{Eq:E}-\eqref{Eq:R} in section \ref{II}. In order to perform path continuation and numerical integration we utilize the asynchronous boundary condition $G(\theta+1,0) = G(\theta,L)$, where $L$ is the length of the cavity. This way also the history of the carrier dynamics evolution is taken into account. For perfectly localized states the gain would completely recover to the equilibrium value in-between two depletions, i.e. $G(\theta+1,0) = G_0$. For numerical simulations of the system Eqs.\eqref{EHaus}-\eqref{QHaus} we utilize the Fourier split step method as outlined in Appendix 4 of \cite{GUR17}.

The two non-local terms in the electric field and the gain equation describe the influence of the cavity geometry. Each pulse experiences a second amplification on its way back through the cavity at $T-2\tau_2 = 2\tau_1$ and hence the gain is depleted twice during one round trip of a single pulse. \\
Utilizing the path continuation package pde2path~\cite{UEC14}, we can reconstruct the branches of both FML and PC solutions and find a good agreement with the previously presented DDE results as exemplary shown for the FML branch in Fig.~\ref{Fig3}(a) (dark blue line). Finding this agreement was also possible when varying the $\alpha$ factors. Furthermore, the lower stability boundary of the FML solution (SN bifurcation) can very well be reproduced by the non-local Haus master equation~\eqref{EHaus}-\eqref{QHaus} as indicated by the dark blue dotted line in Fig.~\ref{Fig5}(a$_1$), which is almost identical to the result obtained from the DDE model \eqref{Eq:E}-\eqref{Eq:R}. \\
Note that in additional to the standard Haus model \cite{HAU00}, the non-local model~\eqref{EHaus}-\eqref{QHaus} makes it possible to study the effects of the cavity geometry combined with the advantages of including dispersion and transverse diffraction effects. A thorough analysis of the differences in the bifurcation structure of the different Haus models are still subject of further investigation, but several differences can be expected as already the first principle DDEs, from which the Haus models can be derived, differ in the stabilizing bifurcations of e.g. HML$_n$ branches (e.g., TR bifurcation instead of SN \cite{MAR14c}, cf. Fig.~\ref{Fig3}). The fundamental branches in both local and non-local cases however appear to be very similar \cite{SCH18e}. Furthermore, we expect differences especially when tuning the value of $\alpha_g$, due to the additional non-locality in the electric field Eq.(\ref{EHaus}). 

\section{BOUND PULSE DISTANCES}
\label{IV}
\begin{figure}[t]
	\includegraphics{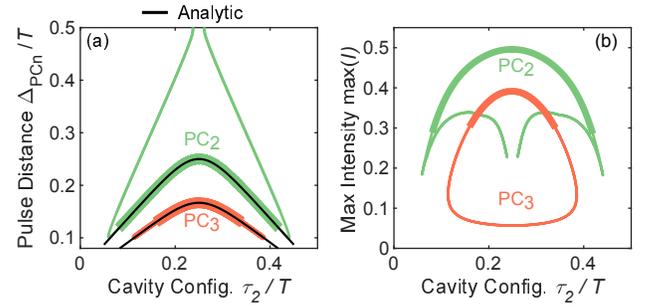}
	\caption{(a) Change of the pulse distances $\Delta_{PC_n}$ within the PC$_2$ (green) and PC$_3$ (red) solution for varied cavity configurations $\tau_2$ calculated utilizing path continuation of the DDEs \eqref{Eq:E}-\eqref{Eq:R} and the analytical expression \eqref{Deltac} (black lines). Thick lines indicate stable regions, thin lines correspond to unstable solutions. (b) Maximum pulse intensity along the branch shown in (a). $G_0/G_{th} = 0.67$, all other parameters as given in Fig.~\ref{Fig2}(a). } 
	\label{Fig6}
\end{figure}

Utilizing the non-local Haus master Eqs.\eqref{EHaus}-\eqref{QHaus} we can derive an analytic expression predicting how the pulse distance $\Delta_{PC_n}$ in regular pulse clusters (see e.g., Fig.~\ref{Fig4}(d$_{1-2}$)) changes with the cavity configuration and how the resulting pulse power depends on the pump power at different cavity configurations. A detailed derivation of the resulting equations for the PC$_2$ solution can be found in Appendix~\ref{APP_Delta}. \\
We start by decomposing the gain and absorber dynamics into slow and fast stage and assume a neglectable pulse width $\varepsilon$. The fast stage in which the gain/absorber is depleted by the pulse can be approximated by: $G_f^{(n)} = G_i^{(n)}e^{-P_n}$ where $G_f^{(n)}$ is the gain value after the pass of pulse $n$, $G_i^{(n)}$ is the gain value just before the incidence of the pulse and $P_n$ is the $n$'th pulse power. On the slow time-scale we approximate the gain relaxation by: 
\begin{align}
G_i^{(n+1)} = G_f^{(n)}e^{-\gamma_g \Delta} + G_0\left[1-e^{-\gamma_g \Delta}\right]\,,
\end{align}
where $\Delta$ is the distance between the pulses and $G_i^{(n+1)}$ is the initial gain value prior to pulse $n+1$. A full set of resulting equations for the decomposition of the gain dynamics, including all depletions, can be found in Appendix~\ref{APP_Delta}. 
In order to be able to solve the system of algebraic equations for $\Delta$, we utilize the extra condition that all pulses are identical, i.e. they experience the same effective gain. Hence, we can recover a relation connecting the pulse energy $P_n = P$ and the distance between the pulses of the PC solution $\Delta$. As the resulting equation is highly singular, a prediction for the pulse distances can be deduced from physical arguments (real and non-negative pulse power - see Fig.\ref{Fig11} in Appendix~\ref{APP_Delta} for details).
This procedure can also be applied for higher order PC solutions. In particular, for the PC$_2$ and PC$_3$ solution the relations read:
\begin{align}
\label{Deltac}
	\Delta_{PC_2} = \frac{T-2\tau_2}{2}-\frac{1}{2\gamma_g } \log\left(\frac{1+e^{\gamma_g(T-4\tau_2)}}{2}\right), \\
		\Delta_{PC_3} = \frac{T-2\tau_2}{3}-\frac{1}{3\gamma_g } \log\left(\frac{1+e^{\gamma_g(T-4\tau_2)}}{2}\right). 
\end{align}
Both distances are only dependent on the cold cavity round-trip time $T$, the gain relaxation $\gamma_g$ and the configuration of the cavity given by $\tau_2$. Utilizing path continuation for the DDE (or PDE) system we obtain the solution branches of the PC$_2$ (PC$_3$) solution in $\tau_2$ (keeping $T$ constant) and determine the pulse distances along the solutions. The result is shown in Fig.~\ref{Fig6}(a), with the path continuation result in green (red) and the respective analytic expressions \eqref{Deltac} in black. 
\ \\
The analytic expression~\eqref{Deltac} accurately matches the path continuation results in the stable regimes, which are depicted by thick lines (thin colored lines indicate unstable solutions). At the outer ends of the stable regimes ($\tau_2 \approx 0.05T$ and $\tau_2 \approx 0.45T$) the PC$_2$ solution branch loops back and the pulse distance grows until the branch reconnects with the HML$_2$ solution at $\tau_2 \approx 0.25T$ (open ends in Fig.~\ref{Fig6}(b)). On the contrary the PC$_3$ solution is a closed loop and reconnects to itself as shown in Fig.~\ref{Fig6}(b) and the pulse distances are equal for the stable and unstable regime. Nevertheless, the amplitudes of stable and unstable solutions are different and drop significantly for the unstable one (Fig.~\ref{Fig6}(b)). This can be explained by the corresponding gain dynamics, which is optimal if the average gain is kept as low as possible (large depletions). In case of the PC$_2$ solution, an unstable regime near the symmetric cavity configuration exists in which the pulses come close to the HML$_2$ branch (top of Fig.~\ref{Fig6}(a)). Here the pulses deplete the gain simultaneously ($\Delta_{PC_2} \rightarrow 0.5T$), which is energetically less favorable and therefore leads to the dynamics being unstable. 
\section{CONCLUSION}
\label{V}
Investigating a passively mode-locked laser with V-shaped external cavity geometry in the long cavity limit, we find that temporal localized structures can be excited individually below the lasing threshold. Due to the V-shaped cavity configuration, in which the gain is depleted twice during each round-trip, phase incoherent molecules can also be stabilized. The latter emerge as clusters of closely packed (incoherent) pulses. Performing a Floquet analysis, we have shown that the pulses within a cluster are globally bound yet locally independent if the cavity is long enough and highlighted that this is a consequence of the non-local interaction with the carriers.
Applying path continuation and direct numerical integration techniques, we predict the stability boundaries of harmonic and pulse cluster solutions with respect to the cavity configuration. This reveals that the stabilization of both types of dynamics is mainly influenced by the maximization of the gain depletion. \\
%Furthermore, we find that the pulse energy and the pulse distance within the PC solutions are additional degrees of freedom, which adapt if the cavity configuration is shifted. 
We have derived a non-local master equation to better understand the non-local influence of the double gain depletion in the V-shaped cavity. Utilizing this non-local PDE system, analytical expressions predicting the pulse power with respect to the gain and most importantly the pulse distance within one cluster with respect to the gain recovery time and the cavity configuration were obtained. In conclusion, we provided for a theoretical framework to tailor the energy and pulse distance in bound pulse clusters in the intermediate and localized regime and facilitate their experimental observation. 
Our findings can have both fundamental and practical interests with regard to the generation of controlled pulse patterns from a single mode-locked laser or the circumvention of unwanted regimes. 
The bound pulse patterns could find applications in optical communications e.g. to implement new bit encoding or at higher powers for material processing. Due to their dense frequency combs, the found clusters could also be utilized in spectroscopy applications requiring a careful distribution of the comb power across several pulses and the timing of arrival at the sample. 

\section{FUNDING}
\label{sec:Funding}
J.H., S. M. and K. L. thank the Deutsche Forschungsgemeinschaft (DFG) within the frame of the SFB787 and the SFB910 for funding. J.J. acknowledge the financial support of the MINECO Project
MOVELIGHT (PGC2018-099637-B-100 AEI/FEDER UE). S.G. acknowledges the
PRIME program of the German Academic Exchange Service (DAAD) with
funds from the German Federal Ministry of Education and Research (BMBF).
\clearpage
\appendix
\section{Coordinate Transformation}
\label{AppTrafo}
In order to give a better insight into the dynamics induced by the cavity geometry, we illustrate the interaction of electric field (pulse) and active sections during one round-trip utilizing the DDE system in which the time is not transformed for computational reasons.
\begin{figure}[b]
	\includegraphics{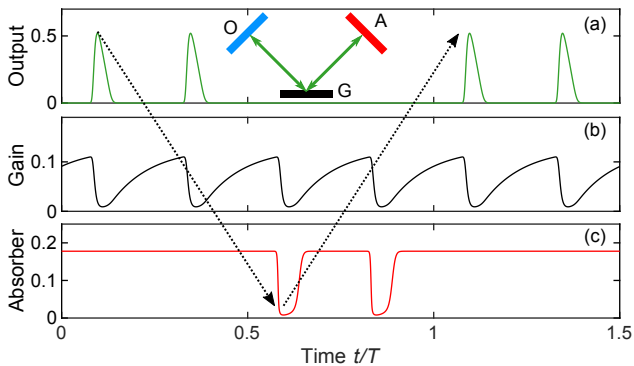}
	\caption{Electric field (green) at the output, gain (black) and absorber (red) dynamics in the untransformed DDE model. The arrow indicates the interaction of the pulse with the active sections during one round-trip, starting and ending at the ouput coupler} 
	\label{Fig12}
\end{figure}
As presented in \cite{HAU19}, this system reads: 
\begin{align}
\label{Eoriginal}
\frac{dE(t^{'})}{dt^{'}} =& -\gamma E(t^{'})+\gamma\sqrt{\kappa} E(t^{'}-2 \tau_ 2 - 2\tau _1)\nonumber \\
& \times e^{\frac{1-i\alpha_g}{2}(\tilde{G}(t^{'}-\tau_1 -2\tau_2)+\tilde{G}(t^{'}-\tau_1))} \nonumber \\
& \times e^{(1-i\alpha_q)\tilde{Q}(t^{'}-\tau_1-\tau_2)}, \\
\label{Goriginal}
\frac{d\tilde{G}(t^{'})}{dt^{'}} =& J_g-\gamma_g\tilde{G}(t^{'})-(e^{\tilde{G}(t^{'})}-1)\nonumber\\
&\times \{\mid E(t^{'}-\tau_1)\mid^2+\mid E(t^{'}-\tau_1-2\tau _2)\mid^2 \nonumber \\
&\times e^{2\tilde{Q}(t^{'}-\tau_2)+\tilde{G}(t^{'}-2\tau_2)}\},\\
\label{Qoriginal}
\frac{d\tilde{Q}(t^{'})}{dt^{'}} =& J_q-\gamma_q \tilde{Q}(t^{'})-r_s(e^{2\tilde{Q}(t^{'})}-1)\nonumber \\ 
&\times e^{\tilde{G}(t^{'}-\tau_2)}\mid E(t^{'}-\tau_1 -\tau_2)\mid^2),
\end{align}
and has several additional delayed terms. The consecutive interaction of a pulse cluster pulse with the active sections in the V-shaped cavity is indicated in Fig.~\ref{Fig12}. Starting at the outcoupling facet, each pulse first depletes the gain after a time $\tau_1$  (first intersection of the dotted arrow with the gain), which is given by the term $\tilde{G}(t-\tau_1-2\tau_2)$ in Eq.~(\ref{Eoriginal}). Hereinafter the pulse depletes the absorber after a time $\tau_1 + \tau_2$ ($\tilde{Q}(t-\tau_1-\tau_2)$) and then again the gain at  $\tau_1 + 2\tau_2$ given by the term  $\tilde{G}(t-\tau_1)$ (second intersection of the dotted arrow with the gain). The pulse cluster results from the equal gain passages, which are energetically favorable. \\
In order to reduce the number of delays to improve the computational costs we introduce the coordinate transformation $t^{'} = t+\tau_1$ for Eq.~(\ref{Goriginal}) and $t^{'} = t+\tau_1+\tau_2$ for Eq.~(\ref{Qoriginal}). We further redefine the variables according to:\\
\begin{align}
\tilde{G}(t^{'}+\tau_1) = G(t),~~~ \tilde{Q}(t^{'}+\tau_1+\tau_2) = Q(t),
\end{align} 
leading to the model presented in Eqs.~(\ref{Eq:E})-(\ref{Eq:R}).

\section{Electrical Triggering of LSs}
\label{AppA}
As demonstrated  experimentally with face-to-face coupled cavities \cite{MAR15b, CAM16}, localized pulses can be triggered via electrical excitation. This can be experimentally achieved by either applying a periodic current modulation \cite{CAM16}, by starting the laser above threshold and then sweeping down the pump current \cite{MAR14c}, or by applying single electrical pulses \cite{CAM16}. 
\begin{figure}[b]
	\includegraphics{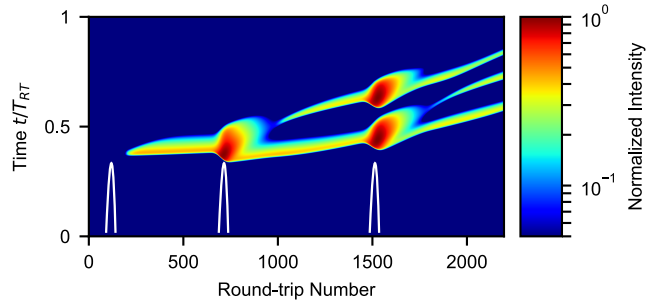}
	\caption{Space-time representation of pulse trains that are excited below the threshold $G_{th}$. The y-axis corresponds to the fast time scale normalized to one period of fundamental solution and the x-axis to the evolution over several periods (approximately round-trips in the cavity). Additional pulses can be written by applying a Gaussian shaped pump pulse to the constant pump $G_0 = 0.85 G_{th}$. The amplitude and FWHM of the first two excitations are $\Delta G_0 = 2.2G_{th}$, $T_{\mathrm{FWHM}} =47T$ and the third one $\Delta G_0 = 2.75G_{th}$, $T_{\mathrm{FWHM}} =59T$. With $\tau_2 = 0.15$ and all other parameters as described in the text. The Gaussian pulse was centered at 100, 700 and 1500 round-trips, indicated by the white lines.} 
	\label{Fig7}
\end{figure}
The resulting dynamics from a numerical investigation of the latter excitation scheme is shown in a pseudo space-time diagram \cite{GIA96} in Fig.~\ref{Fig7}. Here, the y-axis corresponds to the fast time-scale, i.e. changes of the electric field within one round-trip, and the x-axis refers to the evolution from one round-trip to the next. For the electric excitation a Gaussian shaped pump pulse is superimposed to the constant pump current ($G_0 = 0.85 G_{th}$). Three electrical pulses are applied to the system subsequently with a short intermediate transient time, cf. white lines in Fig.~\ref{Fig7}. The amplitudes were $\Delta G_0 = 2.2G_{th}$ for the first two excitation pulses and $\Delta G_0= 2.75G_{th}$ for the third. As can be seen in Fig.~\ref{Fig7}, the first perturbation of the lasing system in the "off" state creates a single localized pulse. Additional pulses can be generated when applying further electrical pulses, as indicated by the second and third excitation. However, in contrast to the case of a face-to-face coupled cavity~\cite{MAR14c, MAR15b, CAM16}, the additional pulse does not stabilize to a harmonic mode-locking solution corresponding to the equidistant pulse spacing. Contrarily, the pulses relax to a state in which they are bound in a cluster by the non-local influence of the second gain depletion within one round-trip, resulting from the V-shaped cavity configuration.

\section{FLOQUET ANALYSIS}
\label{APPFloquet}
The Floquet multipliers \cite{KLA08} give rise to a further classification of how the pulses in one cluster are bound \cite{MAR15d}. They result from the linear stability analysis of the periodic PC solutions of the system. If the absolute value of all Floquet multipliers $\mu$ is less than 1, the solution is considered to be stable. One (trivial) Floquet multiplier can always be found at $\mu = 1$. It corresponds to the neutral mode of the system, when perturbed in the direction that correspond to a translation of the time origin.

\begin{figure}
	\includegraphics{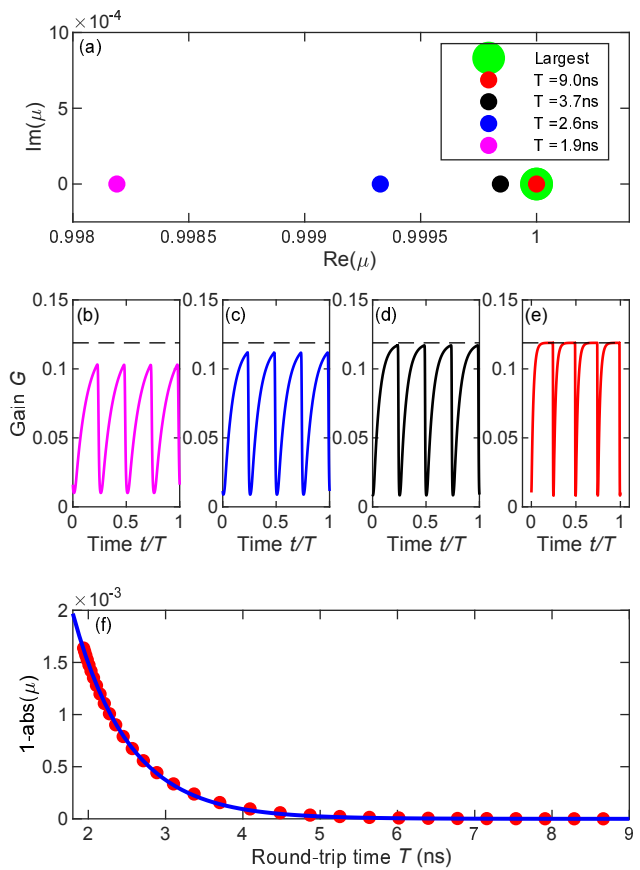}
	\caption{(a) Real and imaginary part of the two maximum Floquet multipliers $\mu$ of the PC$_2$ solutions at different round-trip times are denoted by the colors. At each $T$ the maximum Floquet multiplier is located at $\mu = 1$ indicated by the large green circle. The second largest Floquet multiplier comes closer to $\mu =1$ as the round-trip time is increased. The corresponding gain dynamics is shown in (b)-(e), with the equilibrium value indicated by the dashed line. (f) Distance of the absolute value of the second largest Floquet multiplier to unity at different round-trips found by DDEbiftool (red) and exponential fit (blue). Parameters as in Fig.~\ref{Fig2} and $G_0/G_{th} = 0.65.$} 
	\label{FigFloquet}
\end{figure}

Considering the PC$_2$ solution at high round-trip times, the gain relaxes to its equilibrium value in-between pulses. Therefore, the pulses are locally independent, but globally bound by the depletions induced by the neighboring pulse as shown in Fig.~\ref{FigFloquet}(b). This is supported by the Floquet multipliers, as two neutral modes can be found for this type of pulse clusters, i.e. two Floquet multipliers at $\mu = 1$, marked by the green and red circle in Fig.~\ref{FigFloquet}(a). The corresponding two eigenvectors also refer to the relative temporal translation of either of the two localized states. As the round-trip time is decreased, the gain is not able to relax in-between pulse passes as shown in Fig.~\ref{FigFloquet}(c)-(e), with the equilibrium value indicated by the dashed black line. In this case the second largest Floquet multiplier becomes smaller than unity and therefore the pulses are also locally bound at low $T$, because there is no second neutral-mode (see magenta circle in Fig.~\ref{FigFloquet}(a)). This behavior can also be found for higher order pulse clusters in which the number of neutral modes equals the number of pulses in the cluster, if $T$ is sufficiently high. As the second largest Floquet exponent exponentially approaches unity (shown in Fig.~\ref{FigFloquet}(f)), we also conclude that the transition between locally independent molecules and bound states is continuous depending on the round-trip time. 
\newpage
\section{EVOLUTION OF THE PC$_2$ SOLUTION}
\label{APPPC2}
\begin{figure}[h]
	\includegraphics{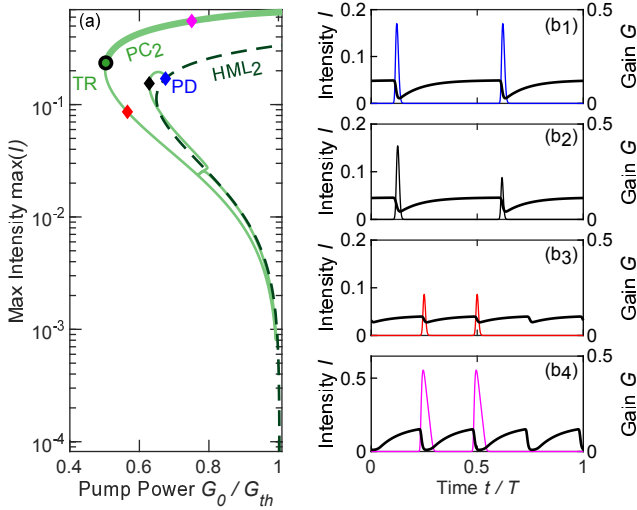}
	\caption{Evolution of the PC$_2$ solution born in a period doubling bifurcation (PD) of the HML$_2$ branch and then stabilizing in a torus bifurcation (TR). (a) Solution branches continued in $G_0$ of the PC$_2$ and HML$_2$ solution. (b$_{1-4}$) Gain and electric field profiles at the points of the branch marked by the colored diamond.} 
	\label{Fig9}
\end{figure}

The localized bound PC$_2$ solution is born unstable in a period doubling bifurcation (PD) along the HML$_2$ branch as indicated by the blue diamond in Fig.\ref{Fig9}(a). As expected the solution is very similar to the HML$_2$ solution close to the bifurcation point (Fig.\ref{Fig9}(b$_1$)). However, along the branch the intensity of both pulses drops subsequently (Fig.\ref{Fig9}(b$_2$)) up to a minimum and as they recover the pulse distance shrinks to $\Delta_{PC_2} = 0.25T$ for $\tau_2 = 0.25T$ (Fig.\ref{Fig9}(b$_3$)) until the pulse cluster regime stabilizes in a torus bifurcation (TR) with corresponding electric field dynamics in Fig.~\ref{Fig9}(b$_4$). \\
As described in the main text, only two gain depletions can be found for the HML$_2$ solution in the symmetric cavity configuration chosen in Fig.~\ref{Fig9}(b$_1$), as the pulse distance $\Delta_{HML_2} = 0.5T$ equals $2\tau_2 = 0.5T$, i.e. two pulses deplete the gain simultaneously. This changes as the pulses come closer from each other in Fig.~\ref{Fig9}(b$_{3-4}$) and deplete the gain at a maximum temporal separation. 

\section{DERIVATION OF THE NON-LOCAL HAUS EQUATION}
\label{APPHaus}
The derivation is based on the multi time-scale approach presented in\cite{KOL06}. We start the derivation with the DDE model \eqref{Eq:E}-\eqref{Eq:R} for the passively mode-locked laser with V-shaped cavity geometry derived in \cite{HAU19}, but rescale the time by one round-trip $t = \frac{\tilde{t}}{T}$. The field is rescaled so that $A = E\gamma_g^{-1}$ and $\tilde{s} = s \frac{\gamma_g}{\gamma_q}$. In contrast to the model equations shown in \cite{HAU19} all other parameters are not rescaled to the round-trip time at this point.
%
% Thus the equations read: 
% %\begin{align}
% \label{Eq:E} \varepsilon \frac{dA}{dt} =& -A+ A(t-1) \sqrt{\kappa}e^{0.5(\tilde{G}(t-\frac{2\tau_1}{T})+\tilde{G}(t-1))-\tilde{Q}(t-1)},\\
% \label{Eq:G}  \frac{d\tilde{G}}{dt} =& \gamma_g T[ G_0-{\tilde{G}}-(e^{\tilde{G}}-1) \nonumber  \\ \times&\{\mid {A}\mid^2+\mid A(t-\frac{2\tau_2}{T})\mid^2e^{-2\tilde{Q}(t-\frac{2\tau_2}{T})+\tilde{G}(t-\frac{2\tau_2}{T})}\}]\\
% \label{Eq:Q} \frac{d\tilde{Q}}{dt}  =& \gamma_q T [Q_0-\tilde{Q}+\tilde{s}(e^{-2\tilde{Q}}-1)e^{\tilde{G}}\mid A\mid^2]
% \end{align}
%
As the localized structures are found in the sub-threshold regime we can assume small gain and absorption.
We introduce the smallness parameter $\varepsilon = \gamma^{-1}T^{-1}$, which is proportional to the pulse width and rescale the variables according to:
\begin{align}
\tilde{G}  = \varepsilon^2 G, \tilde{Q} = \varepsilon^2 Q\\
G_0 = \varepsilon^2 G_m, Q_0 = \varepsilon^2 Q_m\\
\kappa = \exp{-\varepsilon^2 k}
\end{align}
One should note that $\gamma_g T$ is not of the order of $\varepsilon^2$ (in contrast to the short cavity case in \cite{KOL06}), as we investigate the localized regime in which $\gamma T  > 1$ is a necessary condition. Therefore, $\gamma_g T$ and $\gamma_q T$ are not rescaled. \\
Inserting the rescaled variables/parameters and expanding for $\varepsilon$ leads to:
\begin{align}
\varepsilon \frac{dA}{dt} &= -A +A(t-1) \{1 \nonumber \\
&+\frac{\varepsilon^2}{2}[G(t-\frac{2\tau_1}{T})+G(t-1)-2Q(t-1)-k]\}+O(\varepsilon^4),\\
\frac{dG}{dt} &=\gamma_g T [ G_m-G-G\{ |{A}|^2+|A(t-\frac{2\tau_2}{T})|^2\}]+O(\varepsilon^4) \\
\frac{dQ}{dt} &= \gamma_q T [Q_m-Q-2\tilde{s}  Q|A|^2]+O(\varepsilon^2)
\end{align}
We can now introduce a fast time $\sigma$ and a slow time scale $\theta$. We however allow the period to vary on the fast time scale, which is similar to applying the Poincar\'e-Linstedt method. Hence, we define the new time-scales as:
\begin{align}
\sigma &= (1+\omega_1 \varepsilon + \omega_2 \varepsilon^2)t, \\
\theta &= \varepsilon^2 t, 
\end{align}
where the $\omega_j$ will be chosen to make the solvability conditions as simple as possible. Choosing $\mathcal{O}(\varepsilon^2)$ for the slow time-scale is motivated by the time-scale of amplitude (gain) fluctuations and the fast time scale includes period 1 oscillations and the drift induced by the time lag induced by the spectral filter  \cite{VLA05}. The drift can physically be understood by the fact that the period of one pulse train ($T_{PER} \approx T + \gamma^{-1}$) will always be larger than the cold cavity round-trip time $T$, due to causality.
Therefore the total derivative $\frac{d}{dt}$ becomes:
\begin{align}
\frac{d}{dt} \Rightarrow (1+\omega_1 \varepsilon + \omega_2 \varepsilon^2)  \partial_\sigma + \varepsilon^2 \partial_\theta. 
\end{align}
Hence, we find for the delayed terms: 
\begin{align}
&A(\sigma-1-\omega_1\varepsilon-\omega_2\varepsilon^2, \theta-\varepsilon^2) = A(\sigma-1,\theta) \nonumber \\
&-\left(\varepsilon\omega_1+\varepsilon^2\omega_2)\partial_{\sigma}A(\sigma-1,\theta\right)\nonumber \\
&- \varepsilon^2 \left(\partial_{\theta}A(\sigma-1,\theta) +\omega_1^2 \partial^2_{\sigma}A(\sigma-1,\theta)\right)+O(\varepsilon^3)
\end{align}
% \begin{align}
% A(\sigma-1-\omega_1\varepsilon-\omega_2\varepsilon^2, \theta-\varepsilon^2) =& A(\sigma-1,\theta) \nonumber \\
% &-\omega_1 \varepsilon \partial_{\sigma}A(\sigma-1,\theta) \nonumber \\
% &- \omega_2 \varepsilon^2 \partial_{\sigma} A(\sigma-1,\theta) \nonumber \\
% &- \varepsilon^2 \partial_{\theta}A(\sigma-1,\theta)  \nonumber \\
% &- \omega_1^2 \varepsilon^2 \partial^2_{\sigma}A(\sigma-1,\theta) \nonumber \\ &+O(\varepsilon^3)
% \end{align}
We are searching for a solution of the form $A(\sigma, \theta, \varepsilon)$, $G(\sigma, \theta, \varepsilon)$ and $Q(\sigma, \theta, \varepsilon)$; we approximate all variables as a power series in $\varepsilon$:
\begin{align}
	A(t) = A_0(\theta,\sigma)+\varepsilon A_1(\theta,\sigma) + \varepsilon^2 A_2(\theta,\sigma),
\end{align}
and find the equations for the $\mathcal{O}(1)$, $\mathcal{O}(\varepsilon)$ and $\mathcal{O}(\varepsilon^2)$ problems.
Using $A(\sigma, \theta) \widehat{=} A$ and $A(\sigma - 1 , \theta) \widehat{=} A(\sigma -1)$ for simplicity we get for $\mathcal{O}(1)$:
\begin{align}
\label{O(1)}
 A_0 &=A_0(\sigma-1), \\
\partial_{\sigma}G_{0} &= \gamma_g T [ G_m-G_0 \nonumber \\
&-G_0\{ |{A_0}|^2+|A_0(\sigma-\frac{2\tau_2}{T})|^2\}],\\ 
\partial_{\sigma} Q_{0} &= \gamma_q T [Q_m-Q_0-2\tilde{s} Q_0|A_0|^2].
\end{align}
Note that we do not consider the delay in the second variable (the full term that is expanded would read $A(\sigma-\omega \frac{2\tau_1}{T}, \theta - \frac{2\tau_1}{T}\varepsilon^2)$), because it is of order $\varepsilon^2$, which is also true for the higher order terms in $\omega$. Additionally we have neglected the derivative of the slow time-scale, due to the long cavity limit.
From eq.\eqref{O(1)} we can see that $A_0$ is of period 1 . \\
We continue to go to higher order for the electric field and find $\mathcal{O}(\varepsilon)$:
\begin{align}
A_1 -A_1(\sigma-1) &= -\omega_1 \partial_{\sigma}A_{0}(\sigma-1) - \partial_{\sigma} A_{0}  \nonumber \\
\Rightarrow A_1 -A_1(\sigma-1) &= -(1+\omega_1) \partial_{\sigma}A_{0}.
\end{align}
As the $(1+\omega_1)$ term would introduce a resonant forcing, i.e. divergence, we require it to be 0 so that $A_1$ is also period 1. Therefore solvability gives $\omega_1 = -1$ \\
For $\mathcal{O}(\varepsilon^2)$ we get:
% % \begin{align}
% % A_2-A_2(\sigma-1) =&- \omega_1 \partial_{\sigma} A_{0} - \partial_{\sigma} A_{1} \nonumber \\
% % &- \partial_{\theta}A_{0}(\sigma-1)  \nonumber \\
% % &+ \frac{1}{2}\omega_1^2 \partial^2_{\sigma}A_{0}(\sigma_0-1)  \nonumber \\
% % &-  \omega_2 \partial_{\sigma}A_{0}(\sigma-1) \nonumber \\
% % &-  \omega_1 \partial_{\sigma}A_{1}(\sigma-1)\nonumber \\
% % &+\frac{1}{2} [G_0(\sigma-\frac{2\tau_1}{T}) \nonumber \\
% % &+G_0(\sigma-1)+2Q_0(\sigma-1)-k]A_0(\sigma-1).
% % \end{align}
\begin{align}
&A_2-A_2(\sigma-1) =- \omega_1 \partial_{\sigma} A_{0} - \partial_{\sigma} A_{1} - \partial_{\theta}A_{0}(\sigma-1) \nonumber \\
&+ \frac{1}{2}\omega_1^2 \partial^2_{\sigma}A_{0}(\sigma_0-1)-\omega_2 \partial_{\sigma}A_{0}(\sigma-1)-\omega_1 \partial_{\sigma}A_{1}(\sigma-1)\nonumber \\
&+\frac{1}{2}\left[G_0(\sigma-\frac{2\tau_1}{T})+G_0(\sigma-1)+2Q_0(\sigma-1)-k\right]A_0(\sigma-1) .
\end{align}
Using that $A_0$ and $A_1$ are 1 periodic, $\omega_1 = -1$, we find that $\omega_2 = 1$ with the same reasoning as before. Thus, we end up with the following solvability condition:
% \begin{align}
% A_2-A_2(\sigma-1) =& - \partial_{\theta}A_{0} + \frac{1}{2} \partial^2_{\sigma} A_{0}  \nonumber \\
% &+\frac{1}{2} [G_0(\sigma-\frac{2\tau_1}{T})+G_0(\sigma-1) \nonumber \\
% &+2Q_0(\sigma-1)-k]A_0.
% \end{align}
\begin{align}
&A_2-A_2(\sigma-1) = - \partial_{\theta}A_{0} + \frac{1}{2} \partial^2_{\sigma} A_{0}   \\
&+\frac{1}{2}\left[G_0(\sigma-\frac{2\tau_1}{T})+G_0(\sigma-1)+2Q_0(\sigma-1)-k\right]A_0. \nonumber
\end{align}
Note that in the delayed $G_0$ terms the delay for the slow time-scale is neglected as they are of order $\varepsilon^2$ in the expansion, which would give an order of $\varepsilon^4$ in total. \\
The final PDE system then reads:
\begin{align*}
\partial_{\theta}A &= \frac{1}{2} \partial^2_{\sigma}A +\frac{1}{2} [G_0(\sigma-\frac{T-2\tau_2}{T}) \nonumber \\ &+G_0-2Q_0-k]A_0\\
\partial_{\sigma}G_{0} &= \gamma_g T \left[ G_m-G_0-G_0\left\{ |{A_0}|^2+|A_0(\sigma-\frac{2\tau_2}{T})|^2\right\}\right],\\ 
\partial_{\sigma}Q_{0} &= \gamma_q T \left[Q_m-Q_0-2\tilde{s} Q_0|A_0|^2\right]
\end{align*}
Rescaling $\gamma_g$ into $A$, reinserting $\tilde{s}$ and reintroducing $\alpha_g$, $\alpha_q$ finally yields:
\begin{align}
\partial_{\theta} A &=\frac{1}{2 \gamma^2} \partial^2_{\sigma} A +[\frac{1}{2}(1-i\alpha_g) (G(\sigma-\frac{T-2\tau_2}{T})+G)- \nonumber \\ & 
(1-i\alpha_q)Q+\frac{1}{2}\log(\kappa)+i\omega]A,\\
\partial_{\sigma} G &= \gamma_g G_m-\gamma_g G-G\{ |{A}|^2+|A(\sigma-\frac{2\tau_2}{T})|^2\},\\ 
\partial_{\sigma} Q &=  \gamma_q Q_m-\gamma_qQ-2s Q|A|^2.
\end{align}

\section{CARRIER DECOMPOSITION}
\label{APP_Delta}
\begin{figure}[b]
	\includegraphics{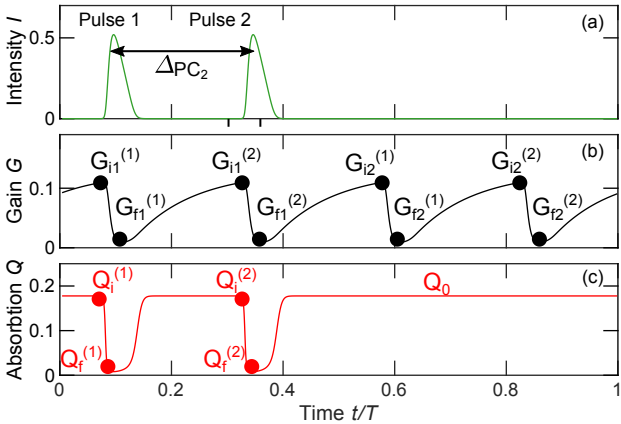}
	\caption{(a) Intensity dynamics of the PC$_2$ solution. (b) Gain depletions of the PC$_2$ solution, with points described in eq.\eqref{G1}-\eqref{G9} marked by black circles. (c) Absorber dynamics, with points described by eq.\eqref{Q1}-\eqref{Q4} marked by red-circles} 
	\label{Fig10}
\end{figure}

In order to find an relation connecting the pulse distance in a pulse cluster $\Delta_{PC_n}$ and the cavity configuration ($T,\tau_1,\tau_2$), we decompose the gain and absorber dynamics into slow and fast stage. We find a sequence of equations for the gain $G_{k{l}}^{(n)}$ and absorber $Q_{k{l}}^{(n)}$, where the subscript $k \in \{i,f\}$ refers to initial and final value of the gain/absorber (before and after the pass of the pulse) and $(n)$ refers to the pulse number. The number $l \in \{i,f\}$ in the subscript indicates the first $1$ or second $2$ gain pass of pulse $(n)$ in one round-trip. The different points are displayed in Fig.~\ref{Fig10}(b-c). For the gain we find the following sequence with pulse energy $P = \int_{-\varepsilon}^{+\varepsilon} |E|^2 d\sigma$, and pulse width $\varepsilon$:\\
\begin{align}
\label{G1}
	t = -\varepsilon\quad,\quad G_{i1}^{(1)},& \\
	t = +\varepsilon\quad,\quad G_{f1}^{(1)} &= G_{i1}^{(1)} e^{-P_1}, \\
    t = \Delta-\varepsilon\quad,\quad G_{i1}^{(2)} &= G_{f1}^{(1)} e^{-\gamma_g \Delta } \nonumber \\
    &+G_0[1-e^{-\gamma_g \Delta}], \\
    t = \Delta+\varepsilon\quad,\quad G_{f1}^{(2)} &= G_{i1}^{(2)} e^{-P_2}, \\
    t = 2\tau_2-\varepsilon\quad,\quad G_{i2}^{(1)} &= G_{f1}^{(2)} e^{-\gamma_g (2\tau_2-\Delta) }\nonumber  \\
    &+G_0[1-e^{-\gamma_g (2\tau_2-\Delta)}], \\
    t = 2\tau_2 +\varepsilon\quad,\quad G_{f1}^{(2)} &= G_{i2}^{(1)} e^{-P_1}, \\
    t = 2\tau_2 +\Delta-\varepsilon\quad,\quad G_{i2}^{(2)} &= G_{f2}^{(1)} e^{-\gamma_g \Delta } \nonumber \\ 
    &+G_0[1-e^{-\gamma_g \Delta}],  \\
    t = 2\tau_2 +\Delta+\varepsilon\quad,\quad G_{f2}^{(2)} &= G_{i2}^{(2)} e^{-P_2}, \\
    t = T \quad,\quad G_{i1}^{(1)} &= G_{f2}^{(2)} e^{-\gamma_g (T-2\tau_2-\Delta) }\nonumber \\
\label{G9}
    &+G_0[1-e^{-\gamma_g (T-2\tau_2-\Delta)}]. 
\end{align}
Due to the fast recovery rate of the absorber, it relaxes to the equilibrium $Q_0$ after each pulse. Furthermore, the absorber is only passed once per round-trip. Therefore the relations read: 
\begin{align}
\label{Q1}
t = -\varepsilon\quad,\quad Q_{i}^{(1)} &= Q_0, \\
t = +\varepsilon\quad,\quad Q_{f}^{(1)} &= Q_0 e^{-2sP_1}, \\
t = \Delta-\varepsilon\quad,\quad Q_{i}^{(2)} &= Q_0, \\
\label{Q4}
t = \Delta+\varepsilon\quad,\quad Q_{f}^{(2)} &= Q_0 e^{-2sP_2}.
\end{align}

\begin{figure}[h]
	\includegraphics{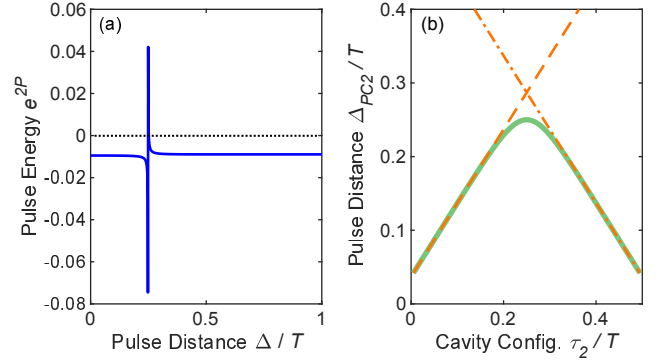}
	\caption{(a) Relationship of pulse power $P$ and pulse distance $\Delta$ given by eq.\eqref{F}, plotted for $T = 1.875$\,ns , $\tau_2 = 0.474T$ and $\gamma_g = 5ns^{-1}$. (b) Solution for the Pulse distance $\Delta_{PC_2}$ for the PC$_2$ solution plotted in green, and asymptotic behavior for $\tau_2 \rightarrow 0,T$ in orange. } 
	\label{Fig11}
\end{figure}

In order to solve the system of the equations for the gain eq.\eqref{G1}-\eqref{G9} for $\Delta$, we utilize the extra condition that the pulses in one cluster are equal, i.e. they experience the same effective gain: \\
\begin{align}
G_{i1}^{(1)} + 	G_{i2}^{(1)} = .. =	G_{i1}^{(n)} +	G_{i2}^{(n)}\,.  
\end{align}
Solving the whole system then leads to the following relation between the pulse power $P$ and the inter pulse distance the pulses $\Delta$:
\begin{align*}
\label{F}
	e^{2P} = \frac{2e^{\gamma_g \Delta} - e^{\gamma_g(2\tau_2 - \Delta)}-e^{\gamma_g (T-\Delta -2\tau_2)}}{2e^{\gamma_g (T-\Delta)} - e^{\gamma_g(2\tau_2 + \Delta)}-e^{\gamma_g (T+\Delta -2\tau_2)}},
\end{align*}
which is only negative, except for the location around the pole $\Delta_{PC_2}$ (see Fig.~\ref{Fig11}(a)). Hence, the physical arguments restrict the solution to be at $\Delta_{PC_2}$, where we find a positive value for $e^{2P}$ and all values of the pulse energy are spanned rapidly. Thus, we can find a relation for the pulse distance, by solving for the pole: 
\begin{align*}
	2e^{\gamma_g (T-\Delta_{PC_2})} - e^{\gamma_g(2\tau_2 + \Delta_{PC_2})}-e^{\gamma_g (T+\Delta_{PC_2} -2\tau_2)} = 0
\end{align*}
which simplifies to:
\begin{align}
		\Delta_{PC_2} = \frac{T-2\tau_2}{2}-\frac{1}{2\gamma_g } \log\left(\frac{1+e^{\gamma_g(T-4\tau_2)}}{2}\right). 
\end{align}
The resulting dependence is visualized Fig\ref{Fig11}(b) for the PC$_2$ solution with the asymptotic for large and small $\tau_2$ shown in orange. These read: \\
\begin{align}
\lim\limits_{\tau_2 \rightarrow 0}\Delta_{PC_2} = \tau_2+\frac{\log(2)}{2\gamma_g}\,, \\
\lim\limits_{2\tau_2 \rightarrow T}\Delta_{PC_2} = \frac{T-2\tau_2}{2}+\frac{\log(2)}{2\gamma_g}. 
\end{align}
In order to also find an expression for the relationship between pulse power and pump current, we further integrate eq.\eqref{EHaus} over each pulse $n$ in the cluster at position $\sigma_n$. We assume the pulse width $\varepsilon$ to be small and apply New's approximation~\cite{NEW74}, which is sufficient in the sub-critical strongly non-linear LS regime, if the out-coupling losses dominate. Furthermore, the pulse energy $P$ does not change on the slow time-scale in a steady state, i.e. $\frac{dP}{d\theta} = 0$. We find that the locus for the pulse existence is given by the zeros of the following integral for the various pulses: 
\begin{align}
\label{EqIj}
\mathcal{I}_n = \int_{\sigma_n-\varepsilon}^{\sigma_n+\varepsilon}[\frac{1}{2}G(\sigma-\tau_1)+\frac{1}{2}G(\sigma)-Q(\sigma)+\frac{1}{2}\log(\kappa) ]|E|^2 d\sigma.
\end{align}
Furthermore, we exploit periodicity $G(\sigma) = G(\sigma +T)$, as we investigate the long cavity regime. As the time-scale of the absorber relaxation is much faster than the gain, it fully recovers between the passing of two pulses (see eq.\eqref{Q1}-\eqref{Q4}). Utilizing these arguments we arrive at the following algebraic expression for the integral in Eq.\eqref{EqIj}:
\begin{align}
\label{EqIj2}
\mathcal{I}_n =& \frac{G_{i1}^{(n)}+G_{i1}^{(n)}}{2}(1-e^{-P_n}) \nonumber \\
&-\frac{Q_0}{2s}(1-e^{-2sP_n})+\frac{1}{2}\log(\kappa)P_n.
\end{align}
Inserting the equations for the gain before the first and second depletion ($G_{i1}^{(n)}$, $G_{i2}^{(n)}$) found from the system of algebraic equations (see eq.\eqref{G1}-\eqref{G9}) and the expression for $\Delta_{PC_2}$ eq.\eqref{Deltac} into eq.\eqref{EqIj2} and solving for the normalized gain $g=\frac{G_0}{G_{th}}$, results in the following equation relating pulse power and the gain value:
\begin{align}
\label{pain}
g(x) =& -4(e^{\gamma_g T}x^2-1)[\frac{Q_0(1-x^{-s})}{2s}-\frac{1}{4}\log(x)\log(\kappa)] \nonumber 
\\ &/ \{(\sqrt{x}-1)(2Q_0 -\log(\kappa))   \nonumber \\ 
&\times [ 2+\frac{\sqrt{2}(\sqrt{x}-1)2e^{\frac{1}{2}\gamma_g(T-2\tau_2)}}{\sqrt{1+e^{\gamma_g(T-4\tau_2)}}} \nonumber \\
&+\frac{\sqrt{2}(\sqrt{x}-1)x e^{\frac{3}{2}\gamma_g(T-2\tau_2)}}{\sqrt{1+e^{\gamma_g(T-4\tau_2)}}} \nonumber \\
&+\frac{\sqrt{2}(\sqrt{x}-1)x e^{\frac{1}{2}\gamma_g(T+2\tau_2)}}{\sqrt{1+e^{\gamma_g(T-4\tau_2)}}} \nonumber \\
&-2e^{\gamma_g T}x^{\frac{3}{2}} \nonumber \\
&+(x-\sqrt{x})(e^{\gamma_g (T-2\tau_2)}+e^{\gamma_g 2\tau_2})
] \} ,
\end{align}
where $x = e^{2P}$ and all other parameters as before.  \\
The characteristics of the folding point of the PC$_2$ solution can also be reproduced by applying the analytical expression (\ref{pain}) for different $\tau_2$ values. Differences might appear due to the fact that the out-coupling losses are very low and hence New's approximation is not perfectly accurate.

\clearpage

\bibliography{export,extra,full}

\end{document}